\documentclass[twocolumn]{jpsj3}
\usepackage{graphicx}
\usepackage{color}
\usepackage{bm}

\title{%
Bulk--Boundary Correspondence and Boundary Zero Modes\\
in a Non-Hermitian Kitaev Chain Model
}

\author{%
Tetsuro Sakaguchi, Hiroto Nishijima, and Yositake Takane
}

\inst{%
Graduate School of Advanced Science and Engineering,\\
Hiroshima University, Higashihiroshima, Hiroshima 739-8530, Japan
}

\recdate{ \hspace{50mm} }

\abst{%
We study a non-Hermitian Kitaev chain model that contains three sources of
non-Hermiticity: a constant imaginary potential,
asymmetry between hopping amplitudes $t_{\rm R}$ and $t_{\rm L}$
in the right and left directions,
and imbalance in pair potentials $\Delta_{\rm c}$ and $\Delta_{\rm a}$
for pair creation and annihilation, respectively.
We show that bulk--boundary correspondence holds in this system;
two topological invariants defined in bulk geometry under a modified
periodic boundary condition correctly describe the presence or absence
of a pair of boundary zero modes in boundary geometry
under an open boundary condition.
One topological invariant characterizes
a topologically nontrivial phase with a line gap
and the other characterizes that with a point gap.
The latter appears
only in the asymmetric hopping case of $t_{\rm R} \neq t_{\rm L}$.
These two nontrivial phases are essentially equivalent
except for their gap structures.
Indeed, the boundary zero modes do not disappear
across the boundary between them.
We also show that the boundary zero modes do not satisfy the Majorana condition
if $\Delta_{\rm c} \neq \Delta_{\rm a}$ and/or $t_{\rm R} \neq t_{\rm L}$.
}

\begin{document}
%%\sloppy
\maketitle

\section{Introduction}

The field of topological materials originates from
studies that elucidate the topological nature of
two-dimensional quantum Hall insulators.~\cite{thouless,kohmoto,hatsugai}
This field has been extended to cover various systems such as
Chern insulators,~\cite{haldane,qi}
topological insulators,~\cite{kane,bernevig,fu1,moore,roy}
and topological superconductors.~\cite{kitaev1,ivanov,fu2,sato1,sato2}
These topological insulators and superconductors are classified
in terms of the ten Altland--Zirnbauer symmetry classes~\cite{altland}
in all spatial dimensions.~\cite{schnyder,kitaev2,ryu1}
Bulk--boundary correspondence is a notable feature
of these topological systems.
Its original scenario~\cite{hatsugai,ryu2} employs
bulk geometry under a periodic boundary condition
and boundary geometry under an open boundary condition.
A topological invariant defined in the bulk geometry predicts the presence or
absence of topological boundary modes in the boundary geometry.

Attempts to extend quantum mechanics to
the non-Hermitian regime~\cite{hatano,bender1,bender2,brody} led to studies
on non-Hermitian topological systems.~\cite{rudner,hu,esaki,ghosh}
A variety of non-Hermitian systems have been considered, including
one-dimensional topological insulators,~\cite{zhu,t_lee,xiong,yao1,alvarez,
yokomizo1,imura1,koch,he,yokomizo2,imura2,kunst1,song,herviou,yuce1}
Chern insulators,~\cite{kunst1,song,herviou,yuce1,yao2,kawabata1,
borgnia,takane1,takane2}
topological semimetals,~\cite{xu2,zyuzin,okugawa,papaj,yokomizo3}
one-dimensional superconductors,~\cite{wang2,yuce2,zeng,klett,menke,li1,
kawabata2,okuma1}
correlated electron systems,~\cite{ashida,yoshida1,yoshida2,e_lee,yoshida3}
Floquet topological systems,~\cite{yuce3,gong1,zhou1,li2,zhou2,
mochizuki1,wu1,li3,bessyo}
and others.~\cite{yuce4,malzard,mochizuki2,leykam,wu2,kondo,kawasaki,
yokomizo4,mostafavi}
Non-Hermitian topological insulators and superconductors are classified
in an exhaustive manner.~\cite{kawabata3}.
The bulk--boundary correspondence is also extensively studied
in non-Hermitian topological systems.
Previous studies showed that it is broken in some cases~\cite{t_lee,xiong}
owing to a non-Hermitian skin effect.~\cite{yao1,longhi1,c_lee,kunst2,
okuma2,zhang,yi,longhi2,kawabata4}
The reason for this is that the non-Hermitian skin effect manifests itself
only in the boundary geometry and vanishes in the bulk geometry.
We are allowed to describe the bulk--boundary correspondence by using
the original scenario with the bulk and boundary geometries
if the non-Hermitian skin effect is absent.
In contrast, we need a special scenario in its presence.
Such scenarios~\cite{yao1,yokomizo1,kunst1,borgnia} employ only
the boundary geometry to avoid the difficulty.

Another scenario employing the bulk and boundary geometries on equal footing
has been proposed in Refs.~\citen{imura1} and \citen{imura2},
where the bulk geometry is defined under a modified periodic boundary condition
[see Eqs.~(\ref{eq:mpbc-R}) and (\ref{eq:mpbc-L})],
which is capable of taking into account the non-Hermitian skin effect.
This scenario is thus applicable
regardless of the presence or absence of the non-Hermitian skin effect.
This has been successfully applied to
one-dimensional topological insulators~\cite{imura1,imura2}
and two-dimensional Chern insulators,~\cite{takane2}
having the potential to describe the bulk--boundary correspondence
in non-Hermitian topological systems in a unified manner.

Let us focus on a Kitaev chain model
for a one-dimensional spinless $p$-wave superconductor.~\cite{kitaev1}
In the topologically nontrivial phase in the Hermitian limit,
this model accommodates a pair of boundary zero modes near its two ends.
The zero modes satisfying the Majorana condition form a nonlocal fermion,
which is a notable feature of this model.
In this paper, we study the bulk--boundary correspondence
in a non-Hermitian Kitaev chain model with three sources of non-Hermiticity:
a constant imaginary potential $i\gamma$ describing gain or loss,
asymmetry between hopping amplitudes $t_{\rm R}$ and $t_{\rm L}$
in the right and left directions,
and imbalance in pair potentials $\Delta_{\rm c}$ and $\Delta_{\rm a}$
for pair creation and annihilation, respectively.
The cases of $\gamma \neq 0$~\cite{zeng,kawabata3} and
$\Delta_{\rm c} \neq \Delta_{\rm a}$~\cite{li1} have been considered
in previous studies.
However, the case of $t_{\rm R} \neq t_{\rm L}$,
as well as the case including all of the three sources, has not been examined.
The purpose of this paper is to show that the scenario in
Refs.~\citen{imura1} and \citen{imura2} correctly describes
the bulk--boundary correspondence in the non-Hermitian Kitaev chain model.
The other purpose is to elucidate how the three sources of non-Hermiticity
affect topological features of the Kitaev chain model.
We show that, in the case of $t_{\rm R} \ne t_{\rm L}$
and/or $\Delta_{\rm c} \neq \Delta_{\rm a}$,
the boundary zero modes do not satisfy the Majorana condition.
Nevertheless, they can form a nonlocal fermion as in the Hermitian limit.
We also show that the nontrivial phase with a point gap appears
in addition to that with a line gap in the case of $t_{\rm R} \neq t_{\rm L}$.

In the next section, we introduce the Hamiltonian of
a non-Hermitian Kitaev chain model that possesses particle--hole symmetry.
We consider eigenvectors of the representation matrix of the Hamiltonian
in the boundary geometry.
By using a biorthogonal set of eigenvectors that reflect
the particle--hole symmetry, we define creation and annihilation operators
for quasiparticles (i.e., bogolons),
in terms of which we can transform the Hamiltonian in a diagonal form.
In Sect.~3, we introduce the bulk geometry
under a modified periodic boundary condition
and give a biorthogonal set of eigenvectors
to define two topological invariants.
In Sect.~4, we apply the scenario of
the bulk--boundary correspondence~\cite{imura1,imura2} to the system.
This gives the condition of Eq.~(\ref{eq:phase-bound}) under which
the topologically nontrivial phase with a pair of boundary zero modes
is realized in the boundary geometry.
In Sect.~5, we confirm the prediction of the bulk--boundary correspondence.
The last section is devoted to a summary and discussion.

\section{Model and Symmetry}

We introduce a Kitaev chain model~\cite{kitaev1} on a one-dimensional lattice
of $N$ sites with lattice constant $a$
and add the three sources of non-Hermiticity to it.
The Hamiltonian is given by
\begin{align}
     \label{eq:def-H}
  H
 & = \sum_{j=1}^{N-1}
     \biggl( -t_{\rm R}c_{j+1}^{\dagger}c_{j} -t_{\rm L}c_{j}^{\dagger}c_{j+1}
       \nonumber \\
 & \hspace{5mm}
          +\Delta_{\rm c}c_{j+1}^{\dagger}c_{j}^{\dagger}
          +\Delta_{\rm a}c_{j}c_{j+1} \biggr)
   - \left(\mu+i\gamma \right) \sum_{j=1}^{N}c_{j}^{\dagger}c_{j} ,
\end{align}
where $c_{j}^{\dagger}$ and $c_{j}$ are respectively the creation and
annihilation operators of the electron at the $j$th site,
and  $t_{\rm R}$, $t_{\rm L}$, $\Delta_{\rm c}$, $\Delta_{\rm a}$, $\mu$,
and $\gamma$ are real parameters.
Here, $t_{\rm R}$ and $t_{\rm L}$ are hopping amplitudes
in the right (positive) and left (negative) directions, respectively,
$\Delta_{\rm c}$ and $\Delta_{\rm a}$ are pair potentials
for $p$-wave pair creation and annihilation, respectively,
$\mu$ is a chemical potential, and $\gamma$ characterizes a constant imaginary
potential describing gain or loss.
We consider only the moderate case of $0 < \Delta_{\rm c}\Delta_{\rm a}$.
The extreme case of $\Delta_{\rm c}\Delta_{\rm a} < 0$~\cite{li1}
may deserve separate consideration.
Equation~(\ref{eq:def-H}) describes the boundary geometry
under the open boundary condition.

We present a theoretical framework to consider the boundary geometry.
This can also apply to the bulk geometry under
the ordinary periodic boundary condition
if we add appropriate boundary terms to $H$.

In terms of column and row vectors defined by
\begin{align}
   C
 & = \,^{t}\!\bigl[c_{1}\, \dots \, c_{N} \,
           c_{1}^{\dagger} \, \dots \, c_{N}^{\dagger}
     \bigr] ,
         \\
   C^{\dagger}
 & = \bigl[c_{1}^{\dagger} \, \dots \, c_{N}^{\dagger} \,
           c_{1} \, \dots \, c_{N}
     \bigr] ,
\end{align}
we rewrite $H$ as $H = \frac{1}{2}C^{\dagger}hC$ with
the $2N \times 2N$ non-Hermitian matrix $h$ given by
\begin{align}
     \label{eq:h-def}
   h =
 \left[
   \begin{array}{cccccccc}
     \hspace{-2mm}-\tilde{\mu} & -t_{\rm L} & 0 & &
           0 & -\Delta_{\rm c} & 0 & \\
     \hspace{-2mm}-t_{\rm R} & -\tilde{\mu} &-t_{\rm L} & &
           \Delta_{\rm c} & 0 & -\Delta_{\rm c} & \\
     \hspace{-2mm}0 & -t_{\rm R} & -\tilde{\mu} & &
           0 & \Delta_{\rm c} & 0 & \\
       & & & \hspace{-3mm}\ddots\hspace{-2mm}
           & & & & \hspace{-3mm}\ddots\hspace{-1mm} \\
     \hspace{-2mm}0 & \Delta_{\rm a} & 0 & &
           \tilde{\mu} & t_{\rm R} & 0 & \\
     \hspace{-2mm}-\Delta_{\rm a} & 0 & \Delta_{\rm a} & &
           t_{\rm L} & \tilde{\mu} & t_{\rm R} & \\
     \hspace{-2mm}0 & -\Delta_{\rm a} & 0 & &
           0 & t_{\rm L} & \tilde{\mu} & \\
       & & & \hspace{-3mm}\ddots\hspace{-2mm}
           & & & & \hspace{-3mm}\ddots\hspace{-1mm}
   \end{array}
  \right] ,
\end{align}
where $\tilde{\mu} = \mu + i \gamma$.
In the Hermitian limit of $\gamma = 0$, $t_{\rm R} = t_{\rm L} \equiv t$,
and $\Delta_{\rm c} = \Delta_{\rm a}$, this model describes
topologically trivial and nontrivial phases.~\cite{kitaev1} 
The nontrivial phase appears when $-2t < \mu < 2t$.
In this phase, a pair of eigenvalues of $h$ become zero.
The zero eigenvalues correspond to a pair of boundary zero modes:
one is localized near the left end of the system
and the other is localized near the right end.~\cite{kitaev1}
The boundary zero modes disappear in the trivial phase.
The topological features described above are preserved
in the non-Hermitian regime, as we describe in Sect.~5.

By using $\tau_{x} = \sigma_{x} \otimes 1_{N \times N}$,
where $\sigma_{x}$ is the $x$-component of Pauli matrices and
$1_{N \times N}$ is the $N \times N$ unit matrix,
we can show that $h$ satisfies
\begin{align}
     \label{eq:PHS}
  \tau_{x}\, ^{t}h\, \tau_{x} = - h ,
\end{align}
which represents the particle-hole symmetry~\cite{kawabata3} of the system.
This symmetry ensures that
a right eigenvector $|\varphi^{R}\rangle$ of $h$ satisfying
\begin{align}
   h|\varphi^{R}\rangle = E|\varphi^{R}\rangle
\end{align}
is paired with a left eigenvector satisfying
\begin{align}
 ^{t}\!\left(\tau_{x}|\varphi^{R}\rangle\right) h
 = \,^{t}\!\left(\tau_{x}|\varphi^{R}\rangle\right)(-E) .
\end{align}
Similarly, a left eigenvector $\langle \varphi^{L}|$ of $h$ satisfying
\begin{align}
   \langle \varphi^{L}|h = \langle \varphi^{L}|E
\end{align}
is paired with a right eigenvector satisfying
\begin{align}
 h\, ^{t}\!\left(\langle \varphi^{L}|\tau_{x}\right)
 = -E\, ^{t}\!\left(\langle \varphi^{L}|\tau_{x}\right) .
\end{align}
These relations ensure that $2N$ eigenvalues of $h$ are written as
$\pm E_{1}$, $\pm E_{2}$, $\dots$, $\pm E_{N}$
with $0 \le \Re\{E_{1}\} \le \Re\{E_{2}\} \le \cdots \le \Re\{E_{N}\}$.
As noted before,
the topologically trivial and nontrivial phases appear in our system.
In the nontrivial phase, one of $\{E_{n}\}$ corresponding to
a pair of boundary zero modes becomes zero.
Without loss of generality, we assume
$E_{1} = 0$ and $E_{n} \neq 0$ for $n = 2,3,\dots, N$,
where $E_{1} = 0$ means that $\Re \{E_{1}\} = \Im \{E_{1}\} = 0$.
In the trivial phase, we assume $E_{n} \neq 0$ for $n = 1,2,\dots, N$.

Let us introduce right and left eigenvectors that satisfy
\begin{align}
   h|\varphi_{n}^{R}\rangle & = -E_{n}|\varphi_{n}^{R}\rangle ,
     \\
   h|\varphi_{N+n}^{R}\rangle & = E_{n}|\varphi_{N+n}^{R}\rangle ,
     \\
   \langle\varphi_{n}^{L}|h & = \langle\varphi_{n}^{L}|\left(-E_{n}\right) ,
     \\
   \langle\varphi_{N+n}^{L}|h & = \langle\varphi_{N+n}^{L}| E_{n} ,
\end{align}
for $n = 1,2,\dots, N$ and
\begin{align}
   \langle \varphi_{n}^{L}|\varphi_{m}^{R}\rangle = \delta_{n,m}
\end{align}
for $n, m = 1,2,\dots, 2N$.
Except in the nontrivial phase,
we can relate the right and left eigenvectors as
\begin{align}
      \label{eq:R-L_1}
  |\varphi_{n}^{R}\rangle &
      = \,^{t}\!\left(\langle\varphi_{N+n}^{L}|\tau_{x}\right) ,
     \\
      \label{eq:R-L_2}
  |\varphi_{N+n}^{R}\rangle &
      = \,^{t}\!\left(\langle\varphi_{n}^{L}|\tau_{x}\right) ,
     \\
      \label{eq:L-R_1}
  \langle\varphi_{n}^{L}| & 
      = \,^{t}\!\left(\tau_{x}|\varphi_{N+n}^{R}\rangle\right) ,
     \\
      \label{eq:L-R_2}
  \langle\varphi_{N+n}^{L}| & 
      = \,^{t}\!\left(\tau_{x}|\varphi_{n}^{R}\rangle\right) ,
\end{align}
for $n = 1,2,\dots, N$.
In the nontrivial phase where
$E_{1} = 0$ and $E_{n} \neq 0$ for $n = 2,3,\dots, N$,
we set the $1$st and $N+1$th eigenvectors with zero eigenvalue
such that
\begin{align}
       \label{eq:L-R_1_zero}
  \langle\varphi_{1}^{L}| & 
      = \,^{t}\!\left(\tau_{x}|\varphi_{1}^{R}\rangle\right) ,
     \\
       \label{eq:L-R_2_zero}
  \langle\varphi_{N+1}^{L}| & 
      = \,^{t}\!\left(\tau_{x}|\varphi_{N+1}^{R}\rangle\right) .
\end{align}
The other eigenvectors obey Eqs.~(\ref{eq:R-L_1})--(\ref{eq:L-R_2}).

To diagonalize $h$, we define the $2N \times 2N$ matrix $V$ as
\begin{align}
      \label{eq:V}
   V = \left[ |\varphi_{1}^{R}\rangle \; \cdots \; |\varphi_{N}^{R}\rangle \;
              |\varphi_{N+1}^{R}\rangle \; \cdots \; |\varphi_{2N}^{R}\rangle
       \right] .
\end{align}
Its inverse matrix $V^{-1}$ is given by
\begin{align}
      \label{eq:V_inv}
   V^{-1} = \left[ \begin{array}{c}
                     \langle\varphi_{1}^{L}| \\
                     \vdots \\
                     \langle\varphi_{N}^{L}| \\
                     \langle\varphi_{N+1}^{L}| \\
                     \vdots \\
                     \langle\varphi_{2N}^{L}|
                  \end{array}
            \right] .
\end{align}
By using $V$ and $V^{-1}$, we can diagonalize $h$ as
\begin{align}
  V^{-1}hV =
  \left[ \begin{array}{cccccc}
            -E_{1} & & & & & \\
              & \hspace{-1mm}\ddots\hspace{-1mm} & & & \bold{0} & \\
              & & -E_{N} & & & \\
              & & & E_{1} & & \\
              & \bold{0} & & & \hspace{-1mm}\ddots\hspace{-1mm} & \\
              & & & & & E_{N}
         \end{array}
  \right] .
\end{align}
Let us define operators that describe quasiparticles in this system as
\begin{align}
       \label{eq:def-d}
     \left[ d_{1} \, \dots \, d_{N} \,
            \bar{d}_{1} \, \dots \, \bar{d}_{N}
     \right]
 & = C^{\dagger}V ,
          \\
       \label{eq:def-f}
     ^{t}\!\left[ \bar{f}_{1} \, \dots \, \bar{f}_{N} \,
                  f_{1} \, \dots \, f_{N}
           \right]
 & = V^{-1}C .
\end{align}
Except in the nontrivial phase, we can show that
\begin{align}
      \label{eq:iden_f-d}
   f_{n} & = d_{n} ,
     \\
     \label{eq:iden_barf-bard}
   \bar{f}_{n} & = \bar{d}_{n}
\end{align}
for $n = 1,2,\dots,N$,
and that $\left\{d_{n}\right\}$ and $\left\{\bar{d}_{n}\right\}$
obey anticommutation relations:
\begin{align}
      \label{eq:commut_1}
 \left\{ d_{n}, \bar{d}_{m}\right\} & = \delta_{n,m} ,
     \\
      \label{eq:commut_2}
 \left\{ d_{n}, d_{m}\right\} & = \left\{ \bar{d}_{n}, \bar{d}_{m}\right\} = 0.
\end{align}
The derivation of Eqs.(\ref{eq:iden_f-d})--(\ref{eq:commut_2})
is outlined in Appendix A.

In the nontrivial phase
where $E_{1} = 0$ and $E_{n} \neq 0$ for $n = 2,3,\dots, N$,
the zero mode operators satisfy
\begin{align}
  f_{1} & = \bar{d}_{1} ,
      \\
  \bar{f}_{1} & = d_{1}
\end{align}
in accordance with Eqs.~(\ref{eq:L-R_1_zero}) and (\ref{eq:L-R_2_zero}).
They also satisfy
\begin{align}
  \left\{ d_{1}, \bar{d}_{n}\right\} & = \left\{ d_{1}, d_{n}\right\} = 0 ,
     \\
  \left\{ \bar{d}_{1}, d_{n}\right\} &
      = \left\{ \bar{d}_{1}, \bar{d}_{n}\right\} = 0
\end{align}
for $n = 2,3,\dots,N$.
Here, $d_{n} = f_{n}$ and $\bar{d}_{n} = \bar{f}_{n}$ ($n = 2,3,\dots,N$)
obey Eqs.~(\ref{eq:commut_1}) and (\ref{eq:commut_2}).
As demonstrated in Sect.~5, the zero mode operators $d_{1}$ and $\bar{d}_{1}$
obey unusual commutation relations [see Eq.~(\ref{eq:commu_zero})].

By using these operators, we can express $H$ as
\begin{align}
      \label{eq:H-bogolon}
   H = \sum_{n=1}^{N}E_{n}\bar{d}_{n}d_{n} ,
\end{align}
where a constant term is subtracted.
In the nontrivial phase, the term with $n = 1$ can be ignored,
or be rewritten as in Eq.~(\ref{eq:H_nontrivial}), since $E_{1} = 0$.
The zero mode operators commute with $H$ as
\begin{align}
  \left[ H, d_{1}  \right] = 0 ,
  \hspace{5mm}
  \left[ H, \bar{d}_{1}  \right] = 0 .
\end{align}
In Sect.~5, $\psi_{1}$ and $\psi_{2}$ defined by
\begin{align}
     \label{eq:def-psi1}
  \psi_{1} & \equiv \sqrt{2} d_{1} ,
      \\
     \label{eq:def-psi2}
  \psi_{2} & \equiv \sqrt{2} \bar{d}_{1}
\end{align}
are used instead of $d_{1}$ and $\bar{d}_{1}$
in the nontrivial phase.

\section{Bulk Geometry}

In this section, we consider the non-Hermitian Kitaev chain model
in the bulk geometry under a modified periodic boundary condition (mpbc).
Let us introduce a pair of plane-wave-like functions given by
\begin{align}
  \varphi_{\beta}^{R}(j) 
   &= \frac{1}{\sqrt{N}}\beta^{j} ,
    \\
  \varphi_{\beta}^{L}(j) 
   &= \frac{1}{\sqrt{N}}\beta^{-j}
\end{align}
with
\begin{align}
  \beta = be^{ika} ,
\end{align}
where $b$ is a real positive constant and
\begin{align}
  k = \frac{2\pi l}{Na}
\end{align}
with $l = 0, 1, \dots, N-1$.
They obey
\begin{align}
    \label{eq:mpbc-R}
  \varphi_{\beta}^{R}(j+N) & = b^{N} \varphi_{\beta}^{R}(j) ,
    \\
    \label{eq:mpbc-L}
  \varphi_{\beta}^{L}(j+N) & = b^{-N} \varphi_{\beta}^{L}(j) ,
\end{align}
which are referred to as the mpbc.

By using the plane-wave-like functions, we define right and left vectors as
\begin{align}
      \label{eq:vR-mpbc}
  |\varphi_{\beta}^{R}\rangle
  & = \sum_{j=1}^{N} \varphi_{\beta}^{R}(j)|j \rangle
      \cdot |\chi_{\beta}^{R}\rangle ,
    \\
      \label{eq:vL-mpbc}
  \langle \varphi_{\beta}^{L}|
  & = \sum_{j=1}^{N} \langle \chi_{\beta}^{L}|
      \cdot \langle j| \varphi_{\beta}^{L}(j) ,
\end{align}
where
\begin{align}
  |\chi_{\beta}^{R} \rangle
   = \left[ \begin{array}{c}
              v_{e}^{R} \\
              v_{h}^{R}
            \end{array}
     \right] ,
    \hspace{6mm}
  \langle \chi_{\beta}^{L}|
   = \left[ \begin{array}{cc}
              \hspace{-1mm} v_{e}^{L} \hspace{-1mm}
              & v_{h}^{L} \hspace{-1mm}
            \end{array}
     \right] ,
\end{align}
and $|j \rangle$ and $\langle j|$ are respectively two-component
row and column vectors:
\begin{align}
  |j \rangle & = \left[ \begin{array}{cc}
                          |j \rangle_{e} & |j \rangle_{h}
                        \end{array}
                 \right] ,
      \\
  \langle j| & = \left[ \begin{array}{c}
                          {}_{e}\langle j| \\
                          {}_{h}\langle j|
                        \end{array}
                 \right] .
\end{align}
Here, $|j \rangle_{e}$ and $|j \rangle_{h}$ are $2N$-component column vectors
and ${}_{e}\langle j|$ and ${}_{h}\langle j|$ are $2N$-component row vectors;
only the $j$th component is $1$ and the others are $0$
in $|j \rangle_{e}$ and ${}_{e}\langle j|$,
and only the $j+N$th component is $1$ and the others are $0$
in $|j \rangle_{h}$ and ${}_{h}\langle j|$.
We define
\begin{align}
  h_{\rm mpbc} = h + \Delta h
\end{align}
such that $|\varphi_{\beta}^{R}\rangle$ and $\langle \varphi_{\beta}^{L}|$
become eigenvectors of $h_{\rm mpbc}$,
where $\Delta h$ consists of eight boundary terms
linking the $1$st and $N$th sites as
\begin{align}
      \label{eq:def-Delta_h}
   \Delta h =
 \left[
   \begin{array}{cccccccc}
     \hspace{-2mm}0 & \hspace{-4mm}\cdots & & \hspace{-1mm}-t_{\rm R}b^{-N} &
        \hspace{-4mm}0 & \hspace{-4mm}\cdots & &
                            \hspace{-1mm}\Delta_{\rm c}b^{-N}\hspace{-2mm} \\
     \hspace{-2mm}\vdots & \hspace{-5mm}\ddots & & \hspace{-1mm} &
        \hspace{-4mm}\vdots & \hspace{-5mm}\ddots & & \hspace{-3mm} \\
     \hspace{-2mm} & & & \hspace{-1mm} &
        \hspace{-4mm} & & & \hspace{-1mm}\hspace{-2mm} \\
     \hspace{-2mm} -t_{\rm L}b^{N} & & & \hspace{-1mm}0 &
        \hspace{-4mm}-\Delta_{\rm c}b^{N} & & & \hspace{-1mm}0\hspace{-2mm} \\
     \hspace{-2mm}0 & \hspace{-4mm}\cdots & &
                            \hspace{-1mm}-\Delta_{\rm a}b^{-N} &
        \hspace{-4mm}0 & \hspace{-4mm}\cdots & &
                            \hspace{-1mm}t_{\rm L}b^{-N}\hspace{-2mm} \\
     \hspace{-2mm}\vdots & \hspace{-5mm}\ddots & & \hspace{-1mm} &
        \hspace{-4mm}\vdots & \hspace{-5mm}\ddots & & \hspace{-3mm} \\
     \hspace{-2mm} & & & \hspace{-1mm} &
        \hspace{-4mm} & & &\hspace{-3mm} \\
     \hspace{-2mm}\Delta_{\rm a}b^{N} & & & \hspace{-1mm}0 &
        \hspace{-4mm}t_{\rm R}b^{N} & & & \hspace{-1mm}0\hspace{-2mm}
   \end{array}
  \right] .
\end{align}
We consider that
the bulk geometry is defined by the representation matrix of $h_{\rm mpbc}$.
Owing to the presence of the boundary terms,
the particle--hole symmetry expressed in Eq.~(\ref{eq:PHS})
holds only when $b = 1$.

The eigenvalue equations of
$h_{\rm mpbc}|\varphi_{\beta}^{R}\rangle = E|\varphi_{\beta}^{R}\rangle$
and $\langle\varphi_{\beta}^{R}|h_{\rm mpbc} = \langle\varphi_{\beta}^{R}|E$
are respectively reduced to
\begin{align}
     \label{eq:red-eve_R}
  h_{\rm rd}(\beta)|\chi_{\beta}^{R}\rangle & = E|\chi_{\beta}^{R}\rangle ,
      \\
     \label{eq:red-eve_L}
  \langle \chi_{\beta}^{L}|h_{\rm rd}(\beta) & = \langle \chi_{\beta}^{L}|E ,
\end{align}
with
\begin{align}
  h_{\rm rd}(\beta)
  = \left[ \begin{array}{cc}
             -t_{\rm R}\beta^{-1}-t_{\rm L}\beta -\tilde{\mu}
             & \Delta_{\rm c}\left(\beta^{-1}-\beta\right) \\
             \Delta_{\rm a}\left(-\beta^{-1}+\beta\right)
             & t_{\rm L}\beta^{-1}+t_{\rm R}\beta +\tilde{\mu}
           \end{array}
    \right] .
\end{align}
It is convenient to define $t$, $\delta t$, and $\bar{\Delta}$ as
\begin{align}
  t & = \frac{t_{\rm R}+t_{\rm L}}{2} ,
      \\
  \delta t & = \frac{t_{\rm R}-t_{\rm L}}{2} ,
      \\
  \bar{\Delta} & = \sqrt{\Delta_{\rm c}\Delta_{\rm a}} ,
\end{align}
where $\bar{\Delta}$ is defined in accordance with the assumption of
$0 < \Delta_{\rm c}\Delta_{\rm a}$.
For simplicity,
we assume $0 < \Delta_{\rm c}$ and $0 < \Delta_{\rm a}$ hereafter.
Solving Eqs.~(\ref{eq:red-eve_R}) and (\ref{eq:red-eve_L}),
we find a pair of eigenvalues of energy:
\begin{align} 
   E_{\pm}(\beta) = \delta t\left(\beta-\beta^{-1} \right)
                    \pm \xi(\beta)
\end{align}
with
\begin{align}
   \xi(\beta)
   = \sqrt{\left(t\left(\beta+\beta^{-1}\right)+\tilde{\mu}\right)^{2}
     - \bar{\Delta}^{2}\left(\beta-\beta^{-1}\right)} ,
\end{align}
where $\xi(\beta)$ is a continuous function of $k$.
The right and left eigenvectors corresponding to $E_{\pm}$ are respectively
given by
\begin{align}
     \label{eq:chi-R-beta}
   |\chi_{\beta\pm}^{R}\rangle
  & = \frac{1}{2}
      \left[ \begin{array}{c}
               \left(\frac{\Delta_{\rm c}}{\Delta_{\rm a}}\right)^{\frac{1}{4}}
               \left(\frac{-t(\beta+\beta^{-1})-\tilde{\mu}
                            +\bar{\Delta}(\beta-\beta^{-1})}{\xi} \pm 1
               \right) \\
               \left(\frac{\Delta_{\rm a}}{\Delta_{\rm c}}\right)^{\frac{1}{4}}
               \left(\frac{-t(\beta+\beta^{-1})-\tilde{\mu}
                            +\bar{\Delta}(\beta-\beta^{-1})}{\xi} \mp 1
               \right)
             \end{array}
      \right] ,
       \\
     \label{eq:chi-L-beta}
   ^{t}\langle\chi_{\beta\pm}^{L}|
  & = \frac{1}{2}
      \left[ \begin{array}{c}
               \left(\frac{\Delta_{\rm a}}{\Delta_{\rm c}}\right)^{\frac{1}{4}}
               \left(\frac{-t(\beta+\beta^{-1})-\tilde{\mu}
                            -\bar{\Delta}(\beta-\beta^{-1})}{\xi} \pm 1
               \right) \\
               \left(\frac{\Delta_{\rm c}}{\Delta_{\rm a}}\right)^{\frac{1}{4}}
               \left(\frac{-t(\beta+\beta^{-1})-\tilde{\mu}
                            -\bar{\Delta}(\beta-\beta^{-1})}{\xi} \mp 1
               \right)
             \end{array}
      \right] .
\end{align}
It is easy to show
\begin{align}
  \langle\chi_{\beta\pm}^{L}|\chi_{\beta\pm}^{R}\rangle = 1 ,
     \hspace{5mm}
  \langle\chi_{\beta\pm}^{L}|\chi_{\beta\mp}^{R}\rangle = 0 .
\end{align}
For $\beta = be^{ika}$ and $\beta' = be^{ik'a}$ with a given $b$,
$\{|\varphi_{\beta}^{R}\rangle\}$ and $\{\langle \varphi_{\beta}^{L}|\}$
constitute a biorthogonal set of eigenvectors as
\begin{align}
  \langle \varphi_{\beta}^{L}|\varphi_{\beta'}^{R}\rangle = \delta_{k,k'} .
\end{align}

\section{Bulk--Boundary Correspondence}

After introducing two topological invariants defined in the bulk geometry,
we examine the bulk--boundary correspondence and specify a condition
that guarantees the appearance of the topologically nontrivial phase
in the boundary geometry.
In this section, in considering the trajectory of $E_{\pm}(\beta)$
with $\beta = be^{ika}$ in the complex energy plane,
we hold $b$ constant and vary $k$ in $[0, \frac{2\pi}{a})$.

Let us assume that a line gap opens between $E_{+}$ and $E_{-}$.
That is, the trajectories of $E_{+}(\beta)$ and $E_{-}(\beta)$
are separated by a line that passes through $E = 0$.
In this case, we define the topological invariant $\nu_{1}$ as
\begin{align}
     \label{eq:def-nu1}
  \nu_{1} = \frac{i}{\pi}\int_{0}^{\frac{2\pi}{a}}dk
            \langle\chi_{\beta-}^{L}|\partial_{k}|\chi_{\beta-}^{R}\rangle .
\end{align}
By using Eqs.~(\ref{eq:chi-R-beta}) and (\ref{eq:chi-L-beta}),
we find
\begin{align}
  \nu_{1} = \frac{\bar{\Delta}}{\pi}\int_{0}^{2\pi}d\theta
            \frac{2t+\frac{\tilde{\mu}}{2}\left(\beta+\beta^{-1}\right)}
                 {\left(t\left(\beta+\beta^{-1}\right)+\tilde{\mu}\right)^{2}
                    - \bar{\Delta}^{2}\left(\beta-\beta^{-1}\right)^{2}} ,
\end{align}
where $\theta = ka$.
In the limit of $b = 1$, this topological invariant becomes equivalent to
that given in Eq.~(96) of Ref.~\citen{kawabata3}
and takes values of $0$ and $1 \pmod 2$:
$\nu_{1} = 0$ corresponds to the topologically trivial phase
and $\nu_{1} = 1$ corresponds to the topologically nontrivial phase
with a pair of boundary zero modes.
Even though $b$ deviates from $b = 1$, the $\mathbb{Z}_{2}$ nature
of $\nu_{1}$ is preserved as long as the line gap opens.
The topological invariant defined in Eq.~(\ref{eq:def-nu1}) is reduced to
the one in Refs.~\citen{sato2} and \citen{ryu1} in the Hermitian limit.

Although $\nu_{1}$ cannot be defined when the line gap is closed,
another topological invariant becomes relevant
if a point gap opens at $E = 0$.
We say that a point gap opens at $E = 0$
if each of $E_{+}(\beta)$ and $E_{-}(\beta)$ forms
a closed loop without passing through $E = 0$
in the absence of the line gap between $E_{+}$ and $E_{-}$
[see Fig. 3(b) as an example of the point gap].
When the point gap opens, we define the topological invariant $\nu_{2}$ as
\begin{align}
   \nu_{2}
   = \frac{1}{4\pi i} \int_{0}^{2\pi}d\theta
     \left[ \frac{d\log E_{+}(\beta)}{d\theta}
            - \frac{d\log E_{-}(\beta)}{d\theta}
     \right] ,
\end{align}
where $\beta = be^{i\theta}$.
In the limit of $b = 1$, this topological invariant also becomes equivalent to
that given in Eq.~(96) of Ref.~\citen{kawabata3}
and takes values of $0$ and $1 \pmod 2$.
Again, $\nu_{2} = 0$ corresponds to the topologically trivial phase
and $\nu_{2} = 1$ corresponds to the topologically nontrivial phase
with a pair of boundary zero modes.
Even though $b$ deviates from $b = 1$,
the $\mathbb{Z}_{2}$ nature of $\nu_{2}$ is preserved
as long as the point gap opens.
The $\mathbb{Z}_{2}$ nature is broken once the point gap is closed
with varying parameters.
Indeed, $\nu_{2}$ can change to $\frac{1}{2}$ (see Fig.~2).
Such a fractional value is irrelevant in considering
the bulk--boundary correspondence.
Note that the sign of $\nu_{2}$ depends on how we determine
the branch of the square root in $\xi(\beta)$.
However, it is also irrelevant
in considering the bulk--boundary correspondence.
Since a point gap appears only in the non-Hermitian regime,
this topological invariant is irrelevant in the Hermitian limit.

Below, we consider $\nu_{1}$ and $\nu_{2}$ in a parameter space
spanned by $\mu$ and $b$.
We first examine the case of $\delta t = 0$
and then turn to the case of $\delta t \neq 0$.

%%%%%%%%%%%%%%%%%%
\begin{figure}[btp]
\begin{center}
\includegraphics[height=4.0cm]{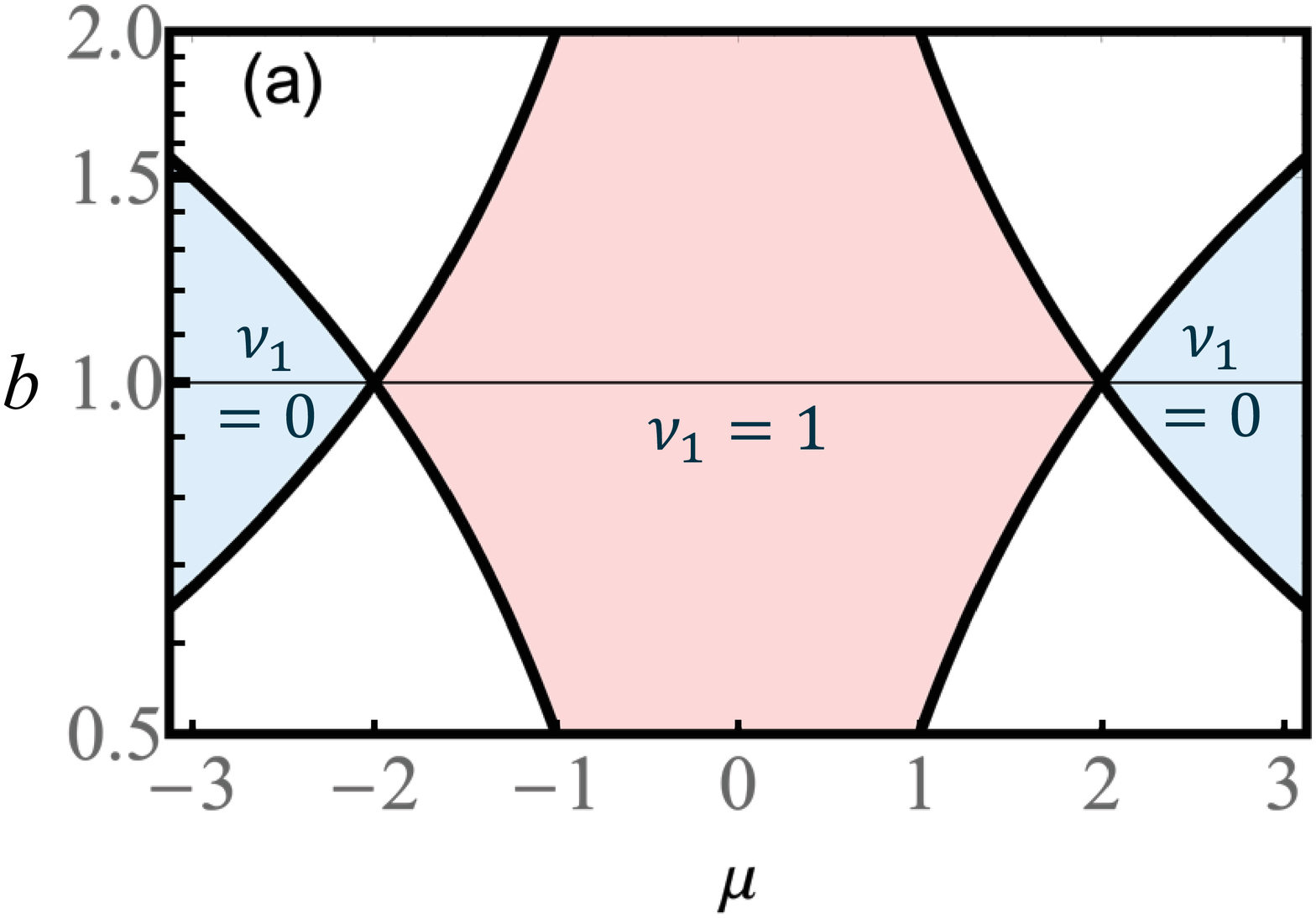}
\includegraphics[height=4.0cm]{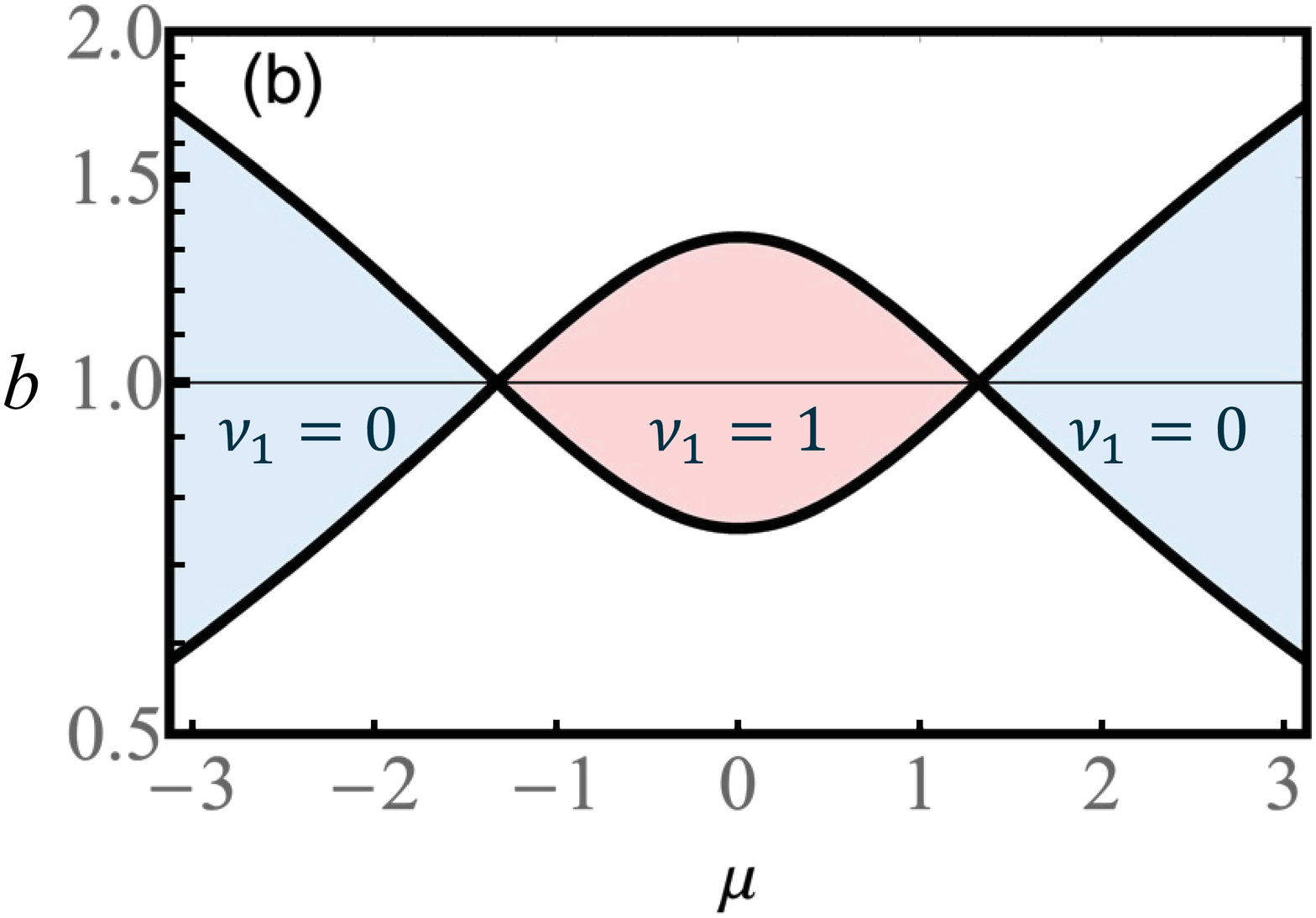}
\end{center}
\caption{
(Color online) Distribution maps of $\nu_{1}$ in the $\mu b$-plane,
where $\delta t/t = 0.0$ and $\bar{\Delta}/t = 1.0$ with
$\gamma/t =$ (a) $0.0$ and (b) $1.5$.
In both maps, the nontrivial region of $\nu_{1} = 1$ (light red) is
in point contact with the two trivial regions of $\nu_{1} = 0$ (light blue).
Thick solid lines represent the gap closing lines of
Eqs.~(\ref{eq:cod-line1}) and (\ref{eq:cod-line2}),
and thin solid lines represent the line of $b = 1$.
}
\end{figure}
%%%%%%%%%%%%%%%%%%
In the case of $\delta t = 0$, where $E_{\pm}(\beta) = \pm\xi(\beta)$,
a gap closing takes place at $E = 0$
when $\xi(\beta) = 0$ for a given $k$.
This indicates that a point gap is forbidden in this case,
and $\nu_{2}$ is irrelevant.
Hence, we need to consider only $\nu_{1}$.
Figure~1 shows the distribution maps of $\nu_{1}$ in the $\mu b$-plane,
where $\delta t/t = 0.0$ and $\bar{\Delta}/t = 1.0$
with $\gamma/t =$ (a) $0.0$ and (b) $1.5$.
The regions of $\nu_{1} = 0$ and $1$ are bounded by lines
on which the line gap closes.
The line gap closes when $\xi(\beta) = 0$ for a given $k$,
resulting in a pair of gap closing lines:
\begin{align}
     \label{eq:cod-line1}
   \frac{\mu^{2}}{\left((t - \bar{\Delta})b
                  + \frac{(t + \bar{\Delta})}{b}\right)^{2}}
 + \frac{\gamma^{2}}{\left((t - \bar{\Delta})b
                     - \frac{(t + \bar{\Delta})}{b}\right)^{2}} = 1 ,
    \\
     \label{eq:cod-line2}
   \frac{\mu^{2}}{\left((t + \bar{\Delta})b
                  + \frac{(t - \bar{\Delta})}{b}\right)^{2}}
 + \frac{\gamma^{2}}{\left((t + \bar{\Delta})b
                     - \frac{(t - \bar{\Delta})}{b}\right)^{2}} = 1 .
\end{align}
Outside the regions of $\nu_{1} = 0$ and $1$,
$\nu_{1}$ cannot be defined because $E_{+}(\beta)$ and $E_{-}(\beta)$ are
combined to form a single band without a gap.
That is, the spectrum becomes gapless
[see Fig.~3(d) as an example of the gapless band].

%%%%%%%%%%%%%%%%%%
\begin{figure}[btp]
\begin{center}
\includegraphics[height=4.1cm]{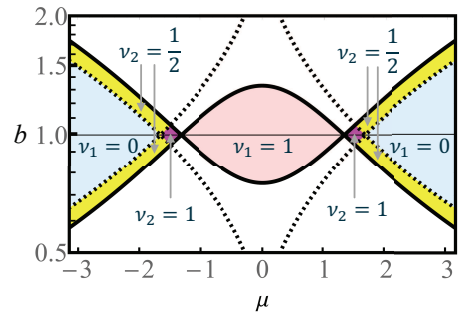}
\end{center}
\caption{
(Color online)  Distribution map of $\nu_{1}$ and $\nu_{2}$
in the $\mu b$-plane, where $\delta t/t = 0.9$ and $\bar{\Delta}/t = 1.0$
with $\gamma/t = 1.5$.
The two nontrivial regions of $\nu_{2} = 1$ (magenta)
appear in this map, where each one is placed between
the nontrivial region of $\nu_{1} = 1$ (light red)
and one of the two trivial regions of $\nu_{1} = 0$ (light blue).
The four elongated regions of $\nu_{2} = \frac{1}{2}$ (yellow) also appear.
Thick solid lines represent the gap closing lines of
Eqs.~(\ref{eq:cod-line1}) and (\ref{eq:cod-line2}),
and dotted lines represent the gap closing lines of
Eqs.~(\ref{eq:cod-point1}) and (\ref{eq:cod-point2}).
}
\end{figure}
%%%%%%%%%%%%%%%%%%
In the case of $\delta t \neq 0$, where
$E_{\pm}(\beta)=\delta t\left(\beta-\beta^{-1} \right)\pm\xi(\beta)$,
a point gap is allowed when $\delta t\left(\beta-\beta^{-1} \right) \neq 0$
at every point in $k \in [0,\frac{2\pi}{a})$ that satisfies $\xi(\beta) = 0$.
That is, a point gap is not forbidden in this case.
Hence, we need to consider $\nu_{1}$ and $\nu_{2}$.
Figure~2 shows the distribution map of $\nu_{1}$ and $\nu_{2}$
in the $\mu b$-plane,
where $\delta t/t = 0.9$ and $\bar{\Delta}/t = 1.0$ with $\gamma/t = 1.5$.
In addition to the gap closing lines of Eqs.~(\ref{eq:cod-line1})
and (\ref{eq:cod-line2}), another pair of gap closing lines for a point gap
are used to separate the regions in Fig.~2.
The point gap closes when
$E_{+}(\beta)=\delta t\left(\beta-\beta^{-1} \right)+\xi(\beta) = 0$
or $E_{-}(\beta)=\delta t\left(\beta-\beta^{-1} \right)-\xi(\beta) = 0$
for a given $k$.
This gives
\begin{align}
     \label{eq:cod-point1}
   \frac{\mu^{2}}{\left((t - \Sigma)b
                  + \frac{(t + \Sigma)}{b}\right)^{2}}
 + \frac{\gamma^{2}}{\left((t - \Sigma)b
                     - \frac{(t + \Sigma)}{b}\right)^{2}} = 1 ,
    \\
     \label{eq:cod-point2}
   \frac{\mu^{2}}{\left((t + \Sigma)b
                  + \frac{(t - \Sigma)}{b}\right)^{2}}
 + \frac{\gamma^{2}}{\left((t + \Sigma)b
                     - \frac{(t - \Sigma)}{b}\right)^{2}} = 1 ,
\end{align}
where
\begin{align}
  \Sigma = \sqrt{\bar{\Delta}^{2}+\delta t^{2}} .
\end{align}
The region of $\nu_{1} = 1$ is bounded by the gap closing lines
of Eqs.~(\ref{eq:cod-line1}) and (\ref{eq:cod-line2}).
The two regions of $\nu_{2} = 1$ are bounded by the gap closing lines
of Eqs.~(\ref{eq:cod-line1})--(\ref{eq:cod-point2}).
The regions of $\nu_{1} = 0$ are bounded by the gap closing lines
of Eqs.~(\ref{eq:cod-point1}) and (\ref{eq:cod-point2}).
These four gap closing lines also separate the regions of
$\nu_{2} = \frac{1}{2}$.
In gapless regions, in which neither $\nu_{1}$ nor $\nu_{2}$
is specified in Fig.~2, both $\nu_{1}$ and $\nu_{2}$ cannot be defined.
Typical spectra in the case of $\delta t \neq 0$ are shown in Fig.~3.
%%%%%%%%%%%%%%%%%%
\begin{figure}[btp]
\begin{center}
\includegraphics[height=3.8cm]{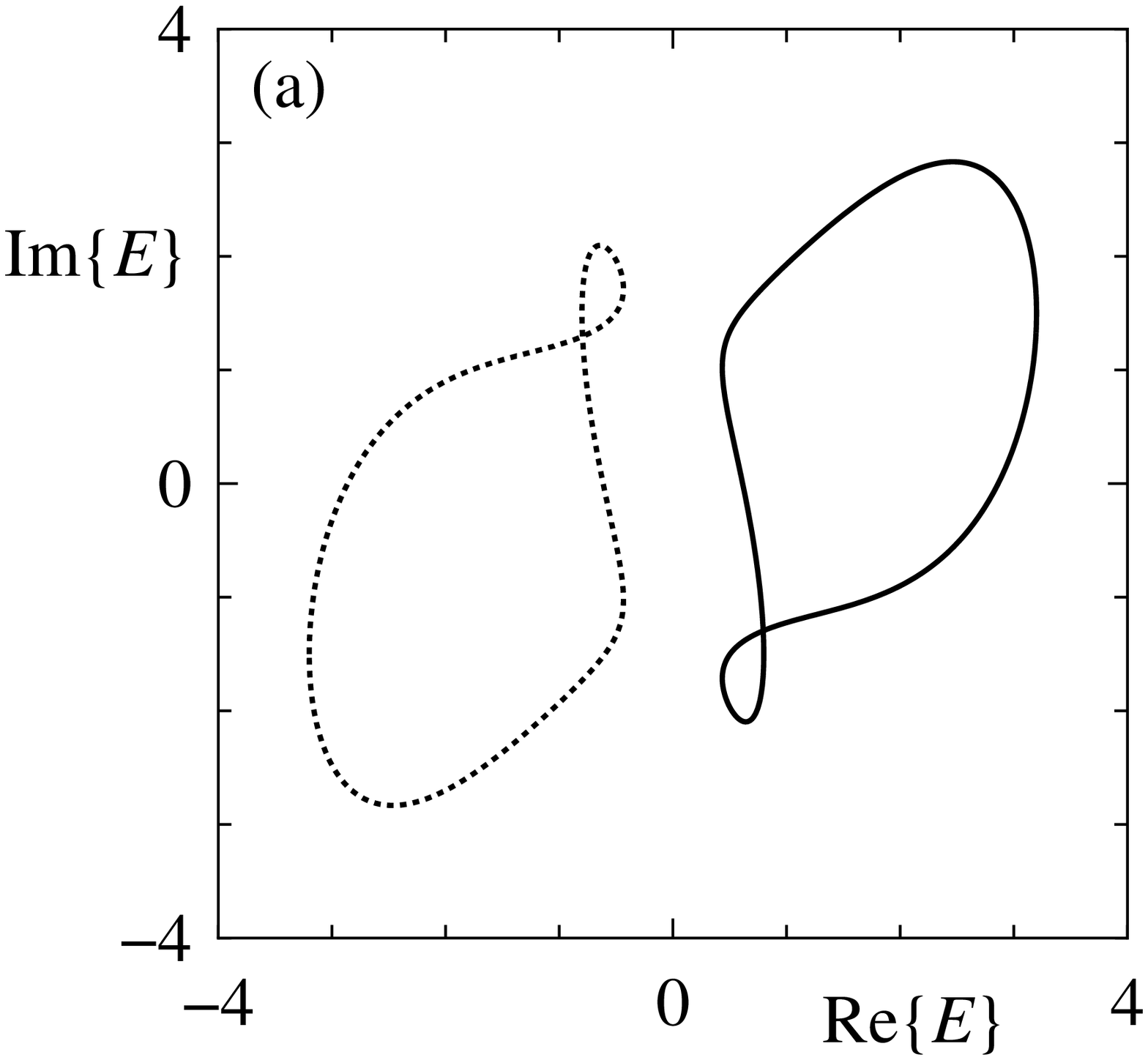}
\includegraphics[height=3.8cm]{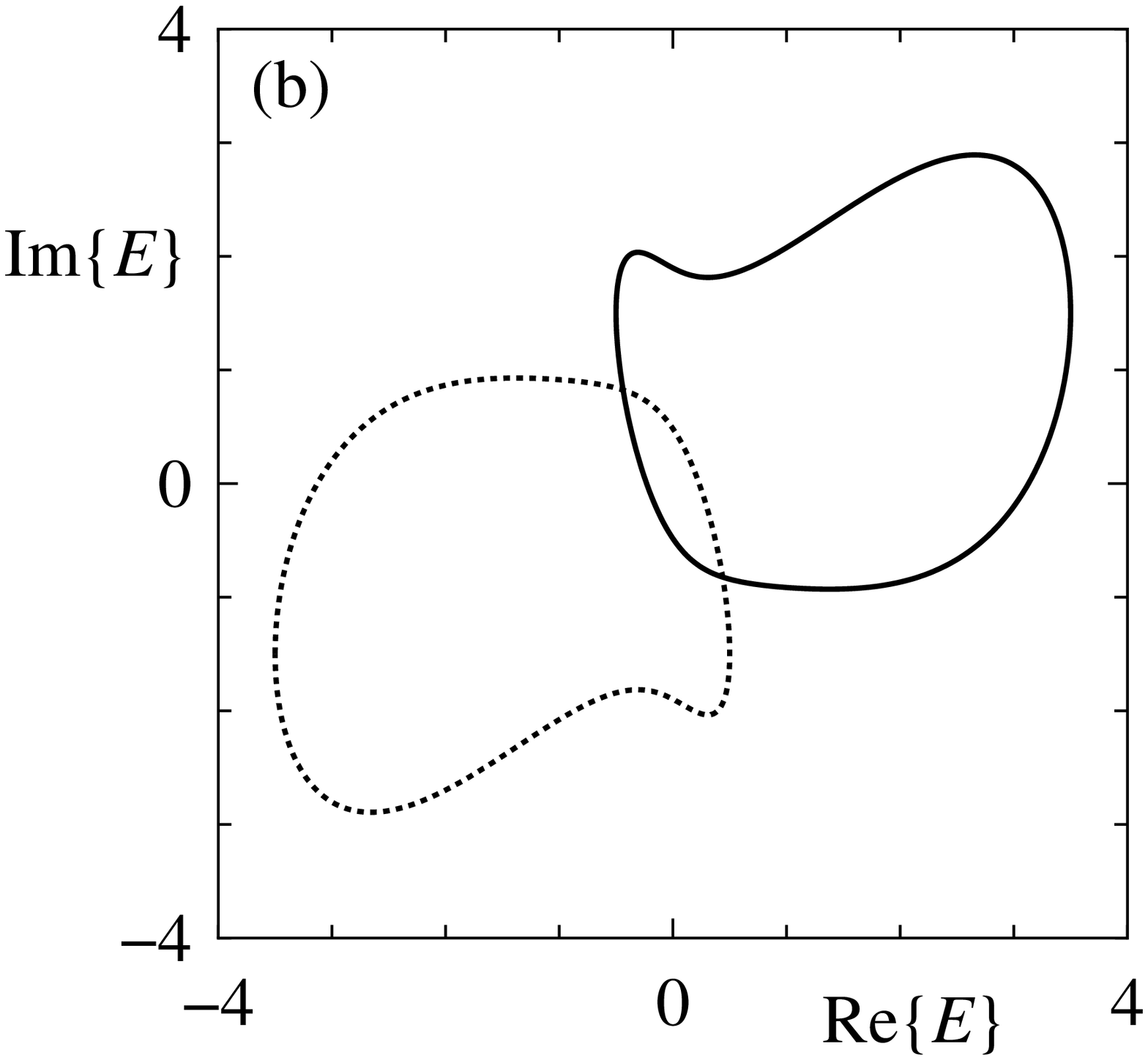}
\includegraphics[height=3.8cm]{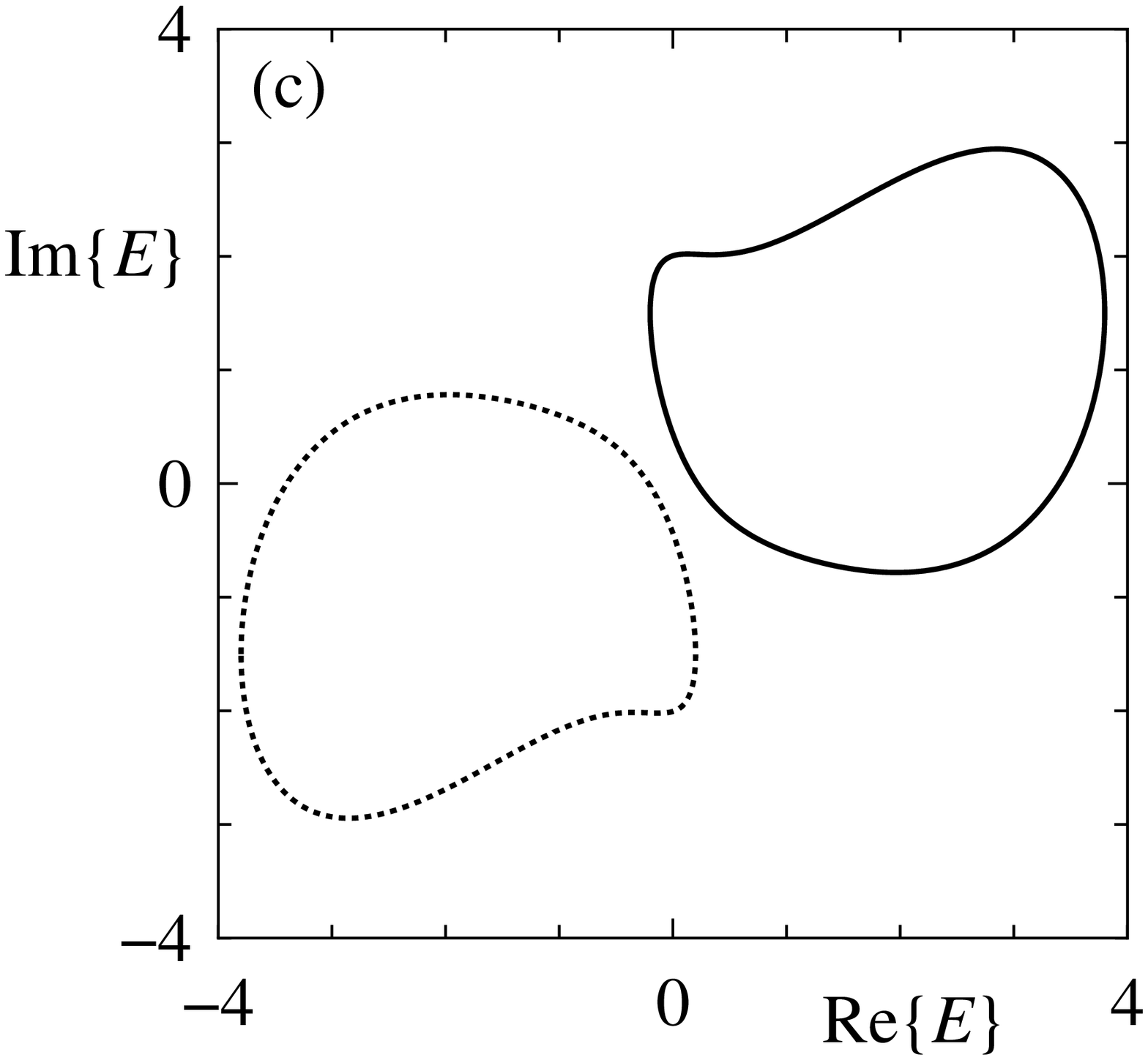}
\includegraphics[height=3.8cm]{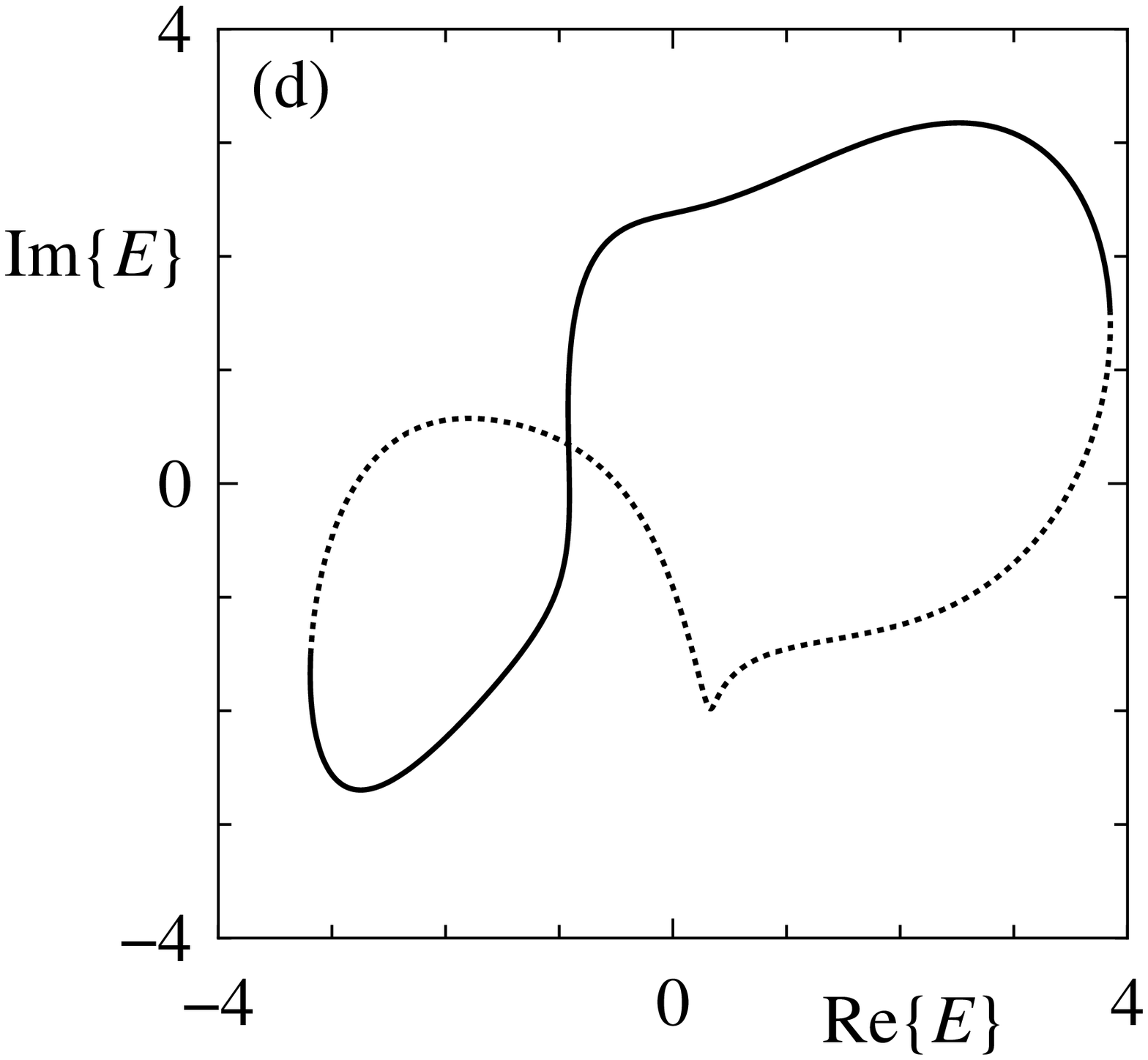}
\end{center}
\caption{
Spectra of the system with $(\mu, b) =$ (a) $(1.2, 1.0)$, (b) $(1.5, 1.0)$,
(c) $(1.8, 1.0)$, and (d) $(1.5, 1.2)$.
Other parameters are
$\delta t/t = 0.9$, $\bar{\Delta}/t = 1.0$, and $\gamma/t = 1.5$.
Solid lines represent $E_{+}(\beta)$
and dotted lines represent $E_{-}(\beta)$.
A line gap opens in (a) and (c), whereas a point gap opens in (b).
In (d), $E_{+}(\beta)$ and $E_{-}(\beta)$ are combined to form a gapless band.
}
\end{figure}
%%%%%%%%%%%%%%%%%%

By using the distribution maps of $\nu_{1}$ and $\nu_{2}$,
let us consider which of the trivial and nontrivial phases appears
in the boundary geometry.
Note that the non-Hermiticity affects $E_{\pm}(\beta)$
through $\gamma$ and $\delta t$.
In the Hermitian limit of $\gamma = 0$ and $\delta t = 0$,
a phase realized in the boundary geometry is governed by $\nu_{1}$
at $b = 1$, which corresponds to the ordinary periodic boundary condition.
If $\nu_{1} = 1$ ($\nu_{1} = 0$) at $b = 1$, the nontrivial (trivial) phase
is realized in the boundary geometry.

To extend this bulk--boundary correspondence to the non-Hermitian regime,
we need to determine $b$ as a function of $\mu$
such that $\nu(\mu,b)$ is in one-to-one correspondence
with a phase realized in the boundary geometry,~\cite{imura1,imura2}
where $\nu(\mu,b)$ represents $\nu_{1}(\mu,b)$ or $\nu_{2}(\mu,b)$
that is well defined for given $\mu$ and $b$.
A recipe for determining $b(\mu)$ is given by adapting the one
in Ref.~\citen{imura2} to our problem.
\begin{enumerate}
\item $b(\mu)$ satisfies $b(\mu) = 1$ at the Hermitian limit and can vary
with increasing $\gamma$ and/or $\delta t$ in a continuous manner.
\item $b(\mu)$ is allowed to cross gap closing lines
only at a crossing point between the two.
\end{enumerate}
The second requirement is based on the following reasoning.~\cite{imura2}
If $b(\mu)$ crosses a gap closing line,
a zero-energy solution appears at the crossing point,
giving rise to a gapless spectrum in the bulk geometry.
Hence, to verify the bulk--boundary correspondence,
the spectrum in the boundary geometry must also be gapless at this point.
A single solution is insufficient to construct a general solution
compatible with the open boundary condition~\cite{yao1,yokomizo1}
that is imposed on the boundary geometry.
A crossing point between two gap closing lines yields
two zero-energy solutions, which should enable us to construct
a general solution at zero energy in the boundary geometry.
Therefore, we expect that $b(\mu)$ is allowed to cross
gap closing lines only at such a crossing point.

Let us apply the recipe to the distribution maps shown in Figs.~1 and 2.
We observe that each pair of gap closing lines
always cross on the line of $b=1$.
Indeed, the gap closing lines of Eqs.~(\ref{eq:cod-line1})
and (\ref{eq:cod-line2}) cross on the line of $b = 1$ at
\begin{align}
    \label{eq:cross-cod-line}
    \mu = \pm 2t\sqrt{1-\left(\frac{\gamma}{2\bar{\Delta}}\right)^{2}} ,
\end{align}
and the gap closing lines of Eqs.~(\ref{eq:cod-point1})
and (\ref{eq:cod-point2}) cross on the line of $b = 1$ at
\begin{align}
    \label{eq:cross-cod-point}
    \mu = \pm 2t\sqrt{1-\left(\frac{\gamma}{2\Sigma}\right)^{2}} .
\end{align}
Equation~(\ref{eq:cross-cod-point}) becomes identical with
Eq.~(\ref{eq:cross-cod-line}) in the limit of $\delta t = 0$.

With these observations, we determine $b(\mu)$ as $b(\mu) = 1$ in all cases
in accordance with the recipe.
Thus, the scenario of Ref.~\citen{imura2} concludes that
$\nu(\mu, b)$ at $b= 1$ is in one-to-one correspondence
with a phase realized in the boundary geometry.
From Figs.~1 and 2, we easily find that the topologically nontrivial phase
is realized in the boundary geometry under the condition of
\begin{align}
    \label{eq:phase-bound}
  -2t\sqrt{1-\left(\frac{\gamma}{2\Sigma}\right)^{2}}
  < \mu <
   2t\sqrt{1-\left(\frac{\gamma}{2\Sigma}\right)^{2}} .
\end{align}
%In the case of $\delta t \ne 0$, Eq.~(\ref{eq:cross-cod-line}) gives
%the boundary between the point gap and line gap regions
%inside the topologically nontrivial phase.

The above argument shows that each gapless point between
the topologically trivial and nontrivial phases is on the line of $b = 1$.
This means that, in the boundary geometry,
a bulk zero-energy state at each gapless point is constructed
by zero-energy solutions at $b = 1$.
That is, such bulk zero-energy states are not subjected to
the non-Hermitian skin effect.

\section{Boundary Geometry}

In this section, we consider the boundary geometry
under the open boundary condition to describe the boundary zero modes.
Assuming that $N$ is sufficiently large, we try to construct
a right eigenvector of $h$ with eigenvalue $E$
by superposing two solutions ($p = \pm$) of
\begin{align}
      \label{eq:eve-localized}
   h_{\rm rd}(\rho_{p})
   \left[ \begin{array}{c}
            u_{p} \\ v_{p}
          \end{array}
   \right]
 = E
   \left[ \begin{array}{c}
            u_{p} \\ v_{p}
          \end{array}
   \right]
\end{align}
as
\begin{align}
      \label{eq:phi-R-localized}
  |\varphi^{R}\rangle
  = \sum_{j=0}^{N+1} |j \rangle\cdot
    \left[ c_{+} \rho_{+}^{j} 
           \left[ \begin{array}{c}
                    u_{+} \\ v_{+}
                  \end{array}
           \right]
         + c_{-} \rho_{-}^{j}
           \left[ \begin{array}{c}
                    u_{-} \\ v_{-}
                  \end{array}
           \right] 
    \right],
\end{align}
where $c_{+}$ and $c_{-}$ are arbitrary constants.
The open boundary condition, which we impose on $|\varphi^{R}\rangle$, is
that the coefficient of $|j \rangle$ vanishes at $j = 0$ and $N+1$.
This is satisfied if $|\rho_{+}| < 1$ and $|\rho_{-}| < 1$ in addition to
\begin{align}
      \label{eq:condition1}
    c_{+}
    \left[ \begin{array}{c}
             u_{+} \\ v_{+}
           \end{array}
    \right]
  + c_{-}
    \left[ \begin{array}{c}
             u_{-} \\ v_{-}
           \end{array}
    \right]
  = \bold{0} .
\end{align}
In this case, $|\varphi^{R}\rangle$ represents an eigenvector
localized near the left end.
The open boundary condition is also satisfied
if $1 < |\rho_{+}|$ and $1 < |\rho_{-}|$ in addition to
\begin{align}
      \label{eq:condition2}
    c_{+} \rho_{+}^{N+1} 
    \left[ \begin{array}{c}
             u_{+} \\ v_{+}
           \end{array}
    \right]
  + c_{-} \rho_{-}^{N+1} 
    \left[ \begin{array}{c}
             u_{-} \\ v_{-}
           \end{array}
    \right]
  = \bold{0} .
\end{align}
In this case, $|\varphi^{R}\rangle$ represents an eigenvector
localized near the right end.

As described in Appendix B, we find two right eigenvectors
with the zero eigenvalue of $E = 0$.
One is localized near the left end
and the other is localized near the right end.
The eigenvector localized near the left end is assigned to
$|\varphi_{1}^{R}\rangle$,
and the eigenvector localized near the right end is assigned to
$|\varphi_{N+1}^{R}\rangle$.
To give their expressions,
let us define $\rho_{1\pm}$ and $\rho_{2\pm}$ as
\begin{align}
      \label{eq:rho1}
  \rho_{1\pm}
  & = \frac{-\tilde{\mu}\pm\sqrt{\tilde{\mu}^{2}
                                 -4\left(t^{2}-\Sigma^{2}\right)}}
           {2\left(t+\Sigma\right)} ,
      \\
      \label{eq:rho2}
  \rho_{2\pm}
  & = \frac{-\tilde{\mu}\pm\sqrt{\tilde{\mu}^{2}
                                 -4\left(t^{2}-\Sigma^{2}\right)}}
           {2\left(t-\Sigma\right)} .
\end{align}
The right eigenvector with $E = 0$ localized near the left end is given by
\begin{align}
    \label{eq:phi-R-left}
  |\varphi_{1}^{R}\rangle
  = c_{R} \sum_{j=1}^{N}
    \left[ \rho_{1+}^{j} - \rho_{1-}^{j}
    \right]
    |j \rangle\cdot
    \left[ \hspace{-1.5mm}
           \begin{array}{c}
             \Delta_{\rm c} \\ \Sigma + \delta t
           \end{array}
           \hspace{-1.5mm}
    \right] ,
\end{align}
where $c_{R}$ is a normalization constant.
This satisfies the open boundary condition when
\begin{align}
      \label{eq:cod-1}
   |\rho_{1+}| < 1, \hspace{4mm} |\rho_{1-}| < 1 .
\end{align}
The corresponding left eigenvector with $E = 0$
localized near the left end is given by
\begin{align}
    \label{eq:phi-L-left}
  \langle \varphi_{1}^{L} |
  = c_{L} \sum_{j=1}^{N}
    \left[ \hspace{-1.5mm}
           \begin{array}{cc}
             \Delta_{\rm a}\hspace{-0.6mm}
             & \hspace{-0.6mm}\Sigma - \delta t
           \end{array}
           \hspace{-1.5mm}
    \right]
    \cdot
    \langle j |
    \left[ \rho_{1+}^{j} - \rho_{1-}^{j} \right] .
\end{align}
The normalization constants are determined as
\begin{align}
  c_{R} = \sqrt{\frac{\Sigma-\delta t}{\Delta_{\rm c}}}
              \frac{c}{\sqrt{2}} ,
  \hspace{6mm}
  c_{L} = \sqrt{\frac{\Delta_{\rm c}}{\Sigma-\delta t}}
              \frac{c}{\sqrt{2}}
\end{align}
with
\begin{align}
  c = \frac{1}{\bar{\Delta}}\left[\sum_{j=1}^{N}
      \left(\rho_{1+}^{j} - \rho_{1-}^{j}\right)^{2}
      \right]^{-\frac{1}{2}}
\end{align}
so that $\langle \varphi_{1}^{L}|\varphi_{1}^{R}\rangle = 1$ and
$\langle\varphi_{1}^{L}|
= \,^{t}\!\left(\tau_{x}|\varphi_{1}^{R}\rangle\right)$.
The right eigenvector with $E = 0$
localized near the right end is given by
\begin{align}
    \label{eq:phi-R-right}
  |\varphi_{N+1}^{R}\rangle
  = c'_{R} \sum_{j=1}^{N}
    \left[ \frac{\rho_{2+}^{j}}{\rho_{2+}^{N+1}}
         - \frac{\rho_{2-}^{j}}{\rho_{2-}^{N+1}}
    \right]
    |j \rangle\cdot
    \left[ \hspace{-1.5mm} 
           \begin{array}{c}
             \Delta_{\rm c} \\ - \left( \Sigma - \delta t \right)
           \end{array}
           \hspace{-1.5mm}
    \right] ,
\end{align}
where $c'_{R}$ is a normalization constant.
This satisfies the open boundary condition when
\begin{align}
      \label{eq:cod-2}
   1 < |\rho_{2+}|, \hspace{4mm} 1 < |\rho_{2-}| .
\end{align}
The corresponding left eigenvector with $E = 0$
localized near the right end is given by
\begin{align}
    \label{eq:phi-L-right}
  \langle \varphi_{N+1}^{L} |
  = c'_{L} \sum_{j=1}^{N}
    \left[ \hspace{-1.5mm}
           \begin{array}{cc}
             -\Delta_{\rm a}\hspace{-0.7mm}
             & \hspace{-0.7mm} \Sigma + \delta t
           \end{array}
           \hspace{-1.5mm}
    \right]
    \cdot \langle j |
    \left[ \frac{\rho_{2+}^{j}}{\rho_{2+}^{N+1}}
         - \frac{\rho_{2-}^{j}}{\rho_{2-}^{N+1}}
    \right] .
\end{align}
The normalization constants are determined as
\begin{align}
  c'_{R} = i\sqrt{\frac{\Sigma+\delta t}{\Delta_{\rm c}}}
                \frac{c'}{\sqrt{2}} ,
  \hspace{6mm}
  c'_{L} = i\sqrt{\frac{\Delta_{\rm c}}{\Sigma+\delta t}}
                \frac{c'}{\sqrt{2}}
\end{align}
with
\begin{align}
  c' = \frac{1}{\bar{\Delta}}
       \left[\sum_{j=1}^{N}
          \left(\frac{\rho_{2+}^{j}}{\rho_{2+}^{N+1}}
                - \frac{\rho_{2-}^{j}}{\rho_{2-}^{N+1}}
          \right)^{2}
       \right]^{-\frac{1}{2}}
\end{align}
so that $\langle \varphi_{N+1}^{L}|\varphi_{N+1}^{R}\rangle = 1$ and
$\langle\varphi_{N+1}^{L}|
=\,^{t}\!\left(\tau_{x}|\varphi_{N+1}^{R}\rangle\right)$.

The right and left eigenvectors with $E = 0$ satisfy
the open boundary condition
when Eqs.~(\ref{eq:cod-1}) and (\ref{eq:cod-2}) hold.
These two equations are equivalent since $\rho_{2\pm} = \rho_{1\mp}^{-1}$,
and are expressed in the simple form
\begin{align}
    \label{eq:cod-1+2}
    |\mu| < 2t\sqrt{1-\left(\frac{\gamma}{2\Sigma}\right)^{2}} .
\end{align}
That is, the boundary zero modes appear in the boundary geometry
if Eq.~(\ref{eq:cod-1+2}) is satisfied.
This is identical with the conclusion of the bulk--boundary correspondence
given in Eq.~(\ref{eq:phase-bound}).

In accordance with Eqs.~(\ref{eq:phi-R-left}) and (\ref{eq:phi-R-right}),
the zero mode operators $\psi_{1}$ and $\psi_{2}$
defined in Eqs.~(\ref{eq:def-psi1}) and (\ref{eq:def-psi2}) are written as
\begin{align}
      \label{eq:zero-mode1}
  \psi_{1}
  & = \sqrt{\frac{\Sigma - \delta t}{\Delta_{\rm c}}}c
      \sum_{j=1}^{N}
      \left[ \rho_{1+}^{j} - \rho_{1-}^{j} \right]
      \left(\Delta_{\rm c}c_{j}^{\dagger}
              + \left(\Sigma + \delta t\right)c_{j}\right) ,
        \\
      \label{eq:zero-mode2}
  \psi_{2}
  & = \sqrt{\frac{\Sigma + \delta t}{\Delta_{\rm c}}}c'
      \sum_{j=1}^{N}
      \left[ \frac{\rho_{2+}^{j}}{\rho_{2+}^{N+1}}
             - \frac{\rho_{2-}^{j}}{\rho_{2-}^{N+1}} \right]
       \nonumber \\
  & \hspace{30mm} \times
      i\left(\Delta_{\rm c}c_{j}^{\dagger}
              - \left(\Sigma - \delta t\right)c_{j}\right) .
\end{align}
Here, $\psi_{1}$ is localized near the left end,
whereas $\psi_{2}$ is localized near the right end.
They commute with $H$ as
$\left[ H, \psi_{1}  \right] = \left[ H, \psi_{2}  \right] = 0$,
and satisfy
\begin{align}
    \label{eq:commu_zero}
  \left\{ \psi_{n}, \psi_{m}\right\} = 2\delta_{n,m} ,
\end{align}
where $n, m = 1,2$.
This anticommutation relation is equivalent to that of Majorana operators,
although $\psi_{1}$ and $\psi_{2}$ cannot be regarded as
pure Majorana operators
in the sense that they do not exactly satisfy the Majorana condition
$\psi_{1}^{\dagger} = \psi_{1}$ and $\psi_{2}^{\dagger} = \psi_{2}$.
In terms of $\psi_{1}$ and $\psi_{2}$, we can form
the annihilation and creation operators of a nonlocal fermion:~\cite{kitaev1}
\begin{align}
  d_{\rm nl} & = \frac{1}{2} \left(\psi_{1} + i\psi_{2}\right) ,
       \\
  \bar{d}_{\rm nl} & = \frac{1}{2}\left(\psi_{1} - i\psi_{2}\right) .
\end{align}
They satisfy anticommutation relations
\begin{align}
  \left\{ d_{\rm nl}, \bar{d}_{\rm nl} \right\} = 1, \hspace{3mm}
  \left\{ d_{\rm nl}, d_{\rm nl} \right\}
  = \left\{ \bar{d}_{\rm nl}, \bar{d}_{\rm nl} \right\} = 0 ,
\end{align}
and commute with $H$ as
$\left[ H, d_{\rm nl}  \right] = \left[ H, \bar{d}_{\rm nl}  \right] = 0$.
It may be instructive to write $H$ as~\cite{li1}
\begin{align}
   \label{eq:H_nontrivial}
   H = 0 \times \bar{d}_{\rm nl}d_{\rm nl}
       + \sum_{n=2}^{N}E_{n}\bar{d}_{n}d_{n} ,
\end{align}
which shows that the ground state is doubly degenerate
because the energy of the system is independent of
whether or not the nonlocal fermion state is occupied.

In the limit of $t_{\rm R} = t_{\rm L}$ and $\Delta_{\rm c}=\Delta_{\rm a}$,
the zero mode operators are simplified to
\begin{align}
  \psi_{1}
  & = c \bar{\Delta}
      \sum_{j=1}^{N}
      \left[ \rho_{1+}^{j} - \rho_{1-}^{j} \right]
      \left(c_{j}^{\dagger} + c_{j}\right) ,
        \\
  \psi_{2}
  & = c' \bar{\Delta}
      \sum_{j=1}^{N}
      \left[ \frac{\rho_{2+}^{j}}{\rho_{2+}^{N+1}}
             - \frac{\rho_{2-}^{j}}{\rho_{2-}^{N+1}} \right]
      i \left(c_{j}^{\dagger} - c_{j}\right) ,
\end{align}
which can be regarded as Majorana operators in the sense that
they are expressed in the form of a linear combination of Majorana operators:
$c_{j}^{\dagger} + c_{j}$ and $i (c_{j}^{\dagger} - c_{j})$.
Equations~(\ref{eq:zero-mode1}) and (\ref{eq:zero-mode2}) show
that the boundary zero modes are modified to unusual forms
if $\Delta_{\rm c} \neq \Delta_{\rm a}$ or $t_{\rm R} \neq t_{\rm L}$.

\section{Summary and Discussion}

We studied the bulk--boundary correspondence
and the characteristics of boundary zero modes
in a non-Hermitian Kitaev chain model that contains
a constant imaginary potential $i\gamma$,
asymmetry between the hopping amplitudes $t_{\rm R}$ and $t_{\rm L}$
in the right and left directions,
and imbalance in pair potentials $\Delta_{\rm c}$ and $\Delta_{\rm a}$
for pair creation and annihilation, respectively.
In the case of $t_{\rm R} \ne t_{\rm L}$ and/or
$\Delta_{\rm c} \neq \Delta_{\rm a}$,
we showed that the boundary zero modes do not satisfy the Majorana condition.
Nevertheless, they can form a nonlocal fermion as in the Hermitian limit.
We also showed that the topologically nontrivial phase with a point gap
appears in addition to that with a line gap
in the case of $t_{\rm R} \neq t_{\rm L}$.

We confirmed that the scenario in Refs.~\citen{imura1} and \citen{imura2}
correctly describes the bulk--boundary correspondence
in the non-Hermitian Kitaev chain model.
In a non-Hermitian superconductor under the open boundary condition,
the non-Hermitian skin effect tends to be suppressed
owing to the coupling between electron and hole sectors.
This is seen from the expression of $h$ in Eq.~(\ref{eq:h-def}).
Indeed, in the electron and hole sectors, the sign of $i\gamma$ is opposite,
and $t_{\rm R}$ and $t_{\rm L}$ are exchanged.
If the non-Hermitian skin effect is strongly suppressed
in low-energy states under the open boundary condition,
we can correctly execute the bulk--boundary correspondence
by using the bulk geometry under the ordinary periodic boundary condition,
as in the Hermitian limit.
That is, the scenario
with the modified periodic boundary condition~\cite{imura1,imura2}
is not indispensable in this case.
Nevertheless, the scenario is still useful because it definitely tells us
whether the non-Hermitian skin effect appears in low-energy states
under the open boundary condition.

\section*{Acknowledgment}

This work was supported by JSPS KAKENHI Grant Number JP21K03405.

\appendix

\section{Derivation of Eqs.~(\ref{eq:iden_f-d})--(\ref{eq:commut_2})}

Equations~(\ref{eq:def-d}) and (\ref{eq:def-f}) give
\begin{align}
  d_{n}
  & = \sum_{j=1}^{N}
      \left(c_{j}^{\dagger}V_{j,n}+c_{j}V_{j+N,n}\right) ,
     \\
  \bar{d}_{n}
  & = \sum_{j=1}^{N}
      \left(c_{j}^{\dagger}V_{j,n+N}+c_{j}V_{j+N,n+N}\right) ,
     \\
  \bar{f}_{n}
  & = \sum_{j=1}^{N}
      \left(V_{n,j}^{-1}c_{j}+V_{n,j+N}^{-1}c_{j}^{\dagger}\right) ,
     \\
  f_{n}
  & = \sum_{j=1}^{N}
      \left(V_{n+N,j}^{-1}c_{j}+V_{n+N,j+N}^{-1}c_{j}^{\dagger}\right) .
\end{align}
From Eqs.~(\ref{eq:R-L_1})--(\ref{eq:L-R_2}), we can show that
\begin{align}
  & V_{n,j}^{-1} = V_{j+N,n+N} ,
     \\
  & V_{n,j+N}^{-1} = V_{j,n+N} ,
     \\
  & V_{n+N,j}^{-1} = V_{j+N,n} ,
     \\
  & V_{n+N,j+N}^{-1} = V_{j,n} .
\end{align}
Combining the eight equations given above,
we immediately obtain Eqs.~(\ref{eq:iden_f-d}) and (\ref{eq:iden_barf-bard}).

Let us turn to the derivation of Eq.~(\ref{eq:commut_1}).
Substituting Eqs.~(\ref{eq:def-d}) and (\ref{eq:def-f}) into
the right-hand side of $\{ d_{n}, \bar{d}_{m} \} = \{ d_{n}, \bar{f}_{m} \}$,
we obtain Eq.~(\ref{eq:commut_1}) as follows:
\begin{align}
 & \{ d_{n}, \bar{d}_{m} \}
           \nonumber \\
 & = \sum_{j=1}^{N}\sum_{k=1}^{N}
     \left\{ c_{j}^{\dagger}V_{j,n}+c_{j}V_{j+N,n},
            V_{m,k}^{-1}c_{k}+V_{m,k+N}^{-1}c_{k}^{\dagger} \right\}
           \nonumber \\
 & = \sum_{j=1}^{N}
     \left(V_{j,n}V_{m,j}^{-1}+V_{j+N,n}V_{m,j+N}^{-1}\right)
   = \delta_{n,m} .
\end{align}
Equation~(\ref{eq:commut_2}) is also derived in a manner similar to this.

\section{Derivation of Eqs.~(\ref{eq:phi-R-left}) and (\ref{eq:phi-R-right})}

We derive two right eigenvectors with $E = 0$
from the eigenvalue equation of Eq.~(\ref{eq:eve-localized}).
One is localized near the right end of the system, whereas the other
is localized near the left end.
The trial function given in Eq.~(\ref{eq:phi-R-localized}) can satisfy
Eq.~(\ref{eq:condition2}), or Eq.~(\ref{eq:condition1}), only if
\begin{align}
   \label{eq:v1=v2}
  ^{t}[u_{+} \hspace{2mm} v_{+}] = \,^{t}[u_{-} \hspace{2mm} v_{-}]
\end{align}
with $\rho_{+} \neq \rho_{-}$,
which results in $E = 0$.~\cite{takane3}
To show this, we rewrite Eq.~(\ref{eq:eve-localized}) as
\begin{align}
     \label{eq:eve-appendix}
   &
   \left[ \begin{array}{cc}
             \frac{-t\left(\rho_{\pm}-\rho_{\pm}^{-1}\right) -\tilde{\mu}-E}
                  {\rho_{\pm}-\rho_{\pm}^{-1}}
             + \delta t
             & -\Delta_{\rm c} \\
             \Delta_{\rm a}
             & \frac{t\left(\rho_{\pm}-\rho_{\pm}^{-1}\right) +\tilde{\mu}-E}
                    {\rho_{\pm}-\rho_{\pm}^{-1}}
               + \delta t
           \end{array}
   \right]
   \left[ \begin{array}{c}
            u_{\pm} \\ v_{\pm}
          \end{array}
   \right]
       \nonumber \\
   & \hspace{10mm}
   = \bold{0} .
\end{align}
Equation~(\ref{eq:v1=v2}) requires that the following two equations,
\begin{align}
  \frac{-t\left(\rho_{+}-\rho_{+}^{-1}\right) -\tilde{\mu}-E}
       {\rho_{+}-\rho_{+}^{-1}}
 & =
  \frac{-t\left(\rho_{-}-\rho_{-}^{-1}\right) -\tilde{\mu}-E}
                  {\rho_{-}-\rho_{-}^{-1}} ,
         \\
  \frac{t\left(\rho_{+}-\rho_{+}^{-1}\right) +\tilde{\mu}-E}
       {\rho_{+}-\rho_{+}^{-1}}
 & =
  \frac{t\left(\rho_{-}-\rho_{-}^{-1}\right) +\tilde{\mu}-E}
       {\rho_{-}-\rho_{-}^{-1}} ,
\end{align}
simultaneously hold.
This straightforwardly results in $E = 0$.

Setting $E = 0$ in Eq.~(\ref{eq:eve-appendix}),
we find that $\rho_{p}$ ($p = \pm$) is determined by
\begin{align}
  t\left(\rho_{p}+\rho_{p}^{-1}\right)+\tilde{\mu}
  = \sigma \Sigma\left(\rho_{p}-\rho_{p}^{-1}\right) ,
\end{align}
where $\sigma = \pm$ and $\Sigma = \sqrt{\bar{\Delta}^{2}+\delta t^{2}}$.
Solving this equation, we find
$\rho_{1\pm}$ given in Eq.~(\ref{eq:rho1}) with
\begin{align}
    \left[ \begin{array}{c}
             u_{1} \\ v_{1}
           \end{array}
    \right]
    =
    \left[ \begin{array}{c}
             \Delta_{\rm c} \\ \Sigma + \delta t
           \end{array}
    \right]
\end{align}
in the case of $\sigma = -$
and $\rho_{2\pm}$ given in Eq.~(\ref{eq:rho2}) with
\begin{align}
    \left[ \begin{array}{c}
             u_{2} \\ v_{2}
           \end{array}
    \right]
    =
    \left[ \begin{array}{c}
             \Delta_{\rm c} \\ -\left( \Sigma - \delta t \right)
           \end{array}
    \right]
\end{align}
in the case of $\sigma = +$.
Substituting $\rho_{1\pm}$ and $^{t}[u_{1} \hspace{2mm} v_{1}]$
into Eq.~(\ref{eq:phi-R-localized}), we find Eq.~(\ref{eq:phi-R-left})
after determining $c_{+}$ and $c_{-}$ in accordance
with Eq.~(\ref{eq:condition1}).
Equation~(\ref{eq:phi-R-right}) is also obtained in a manner similar to this.

\end{document}